%% file: main.tex
\documentclass[10pt,twocolumn,letterpaper]{article}

\usepackage{iccv}
\usepackage{times}
\usepackage{epsfig}
\usepackage{graphicx}
\usepackage{amsmath}
\usepackage{amssymb}
\usepackage{multirow}
\usepackage{booktabs}
\usepackage{gensymb}
\usepackage{color,xcolor}
\usepackage{cite}

\usepackage[pagebackref=true, breaklinks=true,colorlinks,citecolor=citecolor, linkcolor=linkcolor,
letterpaper=true, bookmarks=false]{hyperref}
\definecolor{citecolor}{HTML}{1F801F}
\definecolor{linkcolor}{HTML}{ED1C24}

\usepackage{xspace}
\usepackage{epstopdf}
\usepackage{algorithm}
\usepackage{algorithmic}
\usepackage{diagbox}
\input{math_definition}

\graphicspath{ {./figures/} }

\newlength\savewidth\newcommand\shline{\noalign{\global\savewidth\arrayrulewidth
  \global\arrayrulewidth 1pt}\hline\noalign{\global\arrayrulewidth\savewidth}}
\newcommand{\globalrepresent}{{GloRe}\xspace} 
\newcommand{\methodname}{{GloReDi}\xspace} 
\newcommand{\backbonename}{{Fred-Net}\xspace} 
 
\newcommand{\decoder}{{decoder}\xspace} 
\newcommand{\teacherencoder}{{teacher encoder}\xspace} 
\newcommand{\opname}{\texttt}

\renewcommand{\paragraph}[1]{\vspace{1.25mm}\noindent\textbf{#1}}

\usepackage[capitalize]{cleveref}
\crefname{section}{Sec.}{Secs.}
\Crefname{section}{Section}{Sections}
\Crefname{table}{Table}{Tables}
\crefname{table}{Tab.}{Tabs.}

\iccvfinalcopy % *** Uncomment this line for the final submission

\ificcvfinal\fi

\begin{document}
\title{Learning to Distill Global Representation  for Sparse-View CT}

\author{Zilong Li$^{1}$\qquad Chenglong Ma$^{2}$\qquad Jie Chen$^{1}$\qquad Junping Zhang$^{1}$\qquad Hongming Shan$^{2,3*}$ \\
$^{1}$ Shanghai Key Lab of Intelligent Information Processing, School of Computer Science,\\
Fudan University, Shanghai 200433, China\\
$^{2}$ Institute of Science and Technology for Brain-inspired Intelligence and MOE Frontiers Center \\for Brain Science,  Fudan University, Shanghai 200433, China\\
$^{3}$ Shanghai Center for Brain Science and Brain-inspired Technology, Shanghai 200031, China\\
{\tt\small \{zilongli21, clma22\}@m.fudan.edu.cn,\quad \{chenj19, jpzhang, hmshan\}@fudan.edu.cn} 
}

\maketitle
\ificcvfinal\thispagestyle{empty}\fi

%%%%%%%%% ABSTRACT
\input{./section/abstract.tex}

%%%%%%%%% BODY TEXT
\input{./section/introduction.tex}

\input{./section/related_work_detailed.tex}

\input{./section/method.tex}

\input{./section/result.tex}

\input{./section/conclusion.tex}

\clearpage

{\small
\bibliographystyle{ieee_fullname}
% \bibliography{reference}

}

%%%%%%%%% Appendix
\input{Supplementary.tex}

\end{document}

%% file: math_definition.tex
\usepackage{amsmath,amsfonts}
\usepackage{amsthm}
\usepackage{amssymb,amsopn}
\usepackage{bm} % bold symbol

\newcommand{\vct}[1]{\boldsymbol{#1}} % vector
\newcommand{\mat}[1]{\boldsymbol{#1}} % matrix

\newcommand{\T}{^{\textrm T}} % transpose
\newcommand{\eat}[1]{}

%% file: section/abstract.tex
% !tex root=../main.tex

\begin{abstract}

Sparse-view computed tomography (CT)---using a small number of projections for tomographic reconstruction---enables much lower radiation dose to patients and accelerated data acquisition. The reconstructed images, however, suffer from strong artifacts, greatly limiting their diagnostic value. 
Current trends for sparse-view CT turn to the raw data for better information recovery. The resultant dual-domain methods, nonetheless, suffer from secondary artifacts, especially in ultra-sparse view scenarios, and their generalization to other scanners/protocols is greatly limited. 
A crucial question arises: \emph{have the image post-processing methods reached the limit}? Our answer is not yet. 
In this paper, we stick to image post-processing methods due to great flexibility and propose global representation(\globalrepresent) distillation framework for sparse-view CT, termed \methodname.
First, we propose to learn \globalrepresent with Fourier convolution, so each element in \globalrepresent has an \emph{image-wide} receptive field.
Second, unlike methods that only use the full-view images for supervision, we propose to distill \globalrepresent from intermediate-view reconstructed images that are readily available but not explored in previous literature. 
The success of \globalrepresent distillation is attributed to two key components: representation directional distillation to align the \globalrepresent directions, and band-pass-specific contrastive distillation to gain clinically important details.
Extensive experiments demonstrate the superiority of the proposed \methodname over the state-of-the-art methods, including dual-domain ones.
The source code is available at \url{https://github.com/longzilicart/GloReDi}.
\end{abstract}

%% file: section/introduction.tex
% !tex root=../main.tex

\section{Introduction}\label{sec:intro}

\begin{figure}[t] 
    \centering 
    \includegraphics[width=1\linewidth]{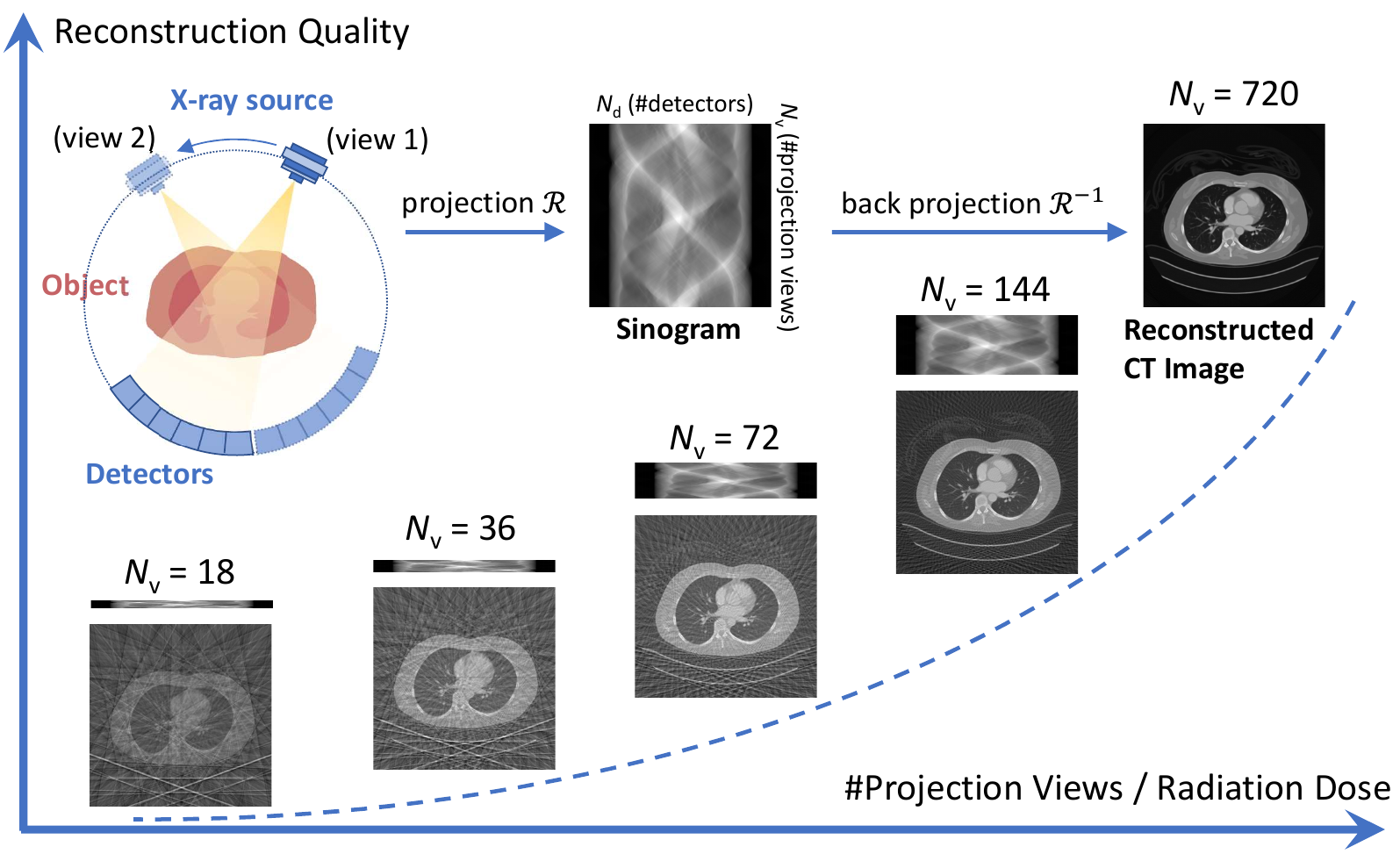}
    \caption{Sparse-view CT only uses a few projection views for tomographic reconstruction, thus providing fast and low-dose CT scanning. However, the reconstruction results of the conventional method suffer from severe \emph{global} streak artifacts. } 
    \label{fig:definition} 
    % \vspace{-2mm}
\end{figure}

X-ray CT is one of the major modalities widely used in clinical screening and diagnosis.
Despite the benefits, there have been growing concerns that X-ray radiation exposure could increase the risk of cancer induction~\cite{ct_outlook}. Following the As Low As Reasonably Achievable (ALARA) principle in the medical community~\cite{alara}, efforts have been made to lower the radiation dose while maintaining imaging quality~\cite{dugan}. Sparse-view CT is one of the effective solutions which reduces the radiation by sampling part of the projection data, \emph{a.k.a} sinogram, for image reconstruction, as shown in Fig.~\ref{fig:definition}. However, analytical reconstruction algorithms such as filtered back projection (FBP) produce inferior image quality with globally severe streak artifacts, significantly compromising its diagnosis value.  
\emph{How to effectively reconstruct the sparse-view CT} remains challenging and, hence, is gaining increasing attention in the computer vision and medical imaging communities.

Image-domain methods regard sparse-view CT reconstruction as an image post-processing task on the FBP-reconstructed images and have achieved exciting performance on streaking artifacts removal and structure preservation~\cite{redcnn, fbpconvnet, ddnet, mwcnn}. However, due to the limited receptive field, these methods have difficulty in modeling the global information efficiently, thus leading to suboptimal results.
Current dominant methods turn to the help of sinogram by restoring the sinogram and performing CT image post-processing simultaneously~\cite{dudonet, hdnet, dudotrans, ddptransformer}. Specifically, in addition to the image domain methods, deep neural networks are also applied to interpolate the missing data in sinogram so that the reconstructed images can be recovered in a global manner~\cite{ssnet, sinogram_complete}.

Despite the common success of dual-domain methods, they suffer from unsatisfactory and unstable performance due to the following reasons. \emph{First}, the processing of sinogram data is sensitive and even subtle changes may remarkably affect the reconstructed images and introduce stubborn secondary artifacts that are hard to be eliminated. 
\emph{Second}, in ultra sparse-view scenarios, inpainting methods are unable to accurately restore the excessive missing raw data; \eg, inpainting sparse CT data of 18 views to the full CT data of 720 views only retains about 2.5\% of the valid information, in which the involvement of sinogram processing makes the learning more difficult and compromise the performance~\cite{dudotrans,ddptransformer}.
\emph{Third}, it is often impractical to access sinogram data given the privacy and commercial concerns, and if accessible, the requirement of raw data greatly limits the generalization to other scanners/protocols.

\paragraph{Motivation.} Given the great flexibility of the image domain methods, we arise a critical question in this paper: \emph{have the image domain (post-processing) methods reached the limits?} Our answer is not yet, given the following observations.
\emph{First}, the existing image-domain methods typically suffer from \emph{limited} receptive field, which fails to extract and recover the global information efficiently~\cite{lama}, given that streak artifacts are spread globally on the reconstructed images as shown in Fig.~\ref{fig:definition}. 
\emph{Second}, the existing image-domain methods typically use the full-view images as the unique supervision to improve the image quality in an end-to-end manner~\cite{ddnet,fbpconvnet}. They ignore the importance of representation learning, especially global representation (\globalrepresent), making the artifact removal and detail recovery entangled, and leading to sub-optimal results due to the significant gap between sparse- and full-view.
\emph{Third}, the intermediate-view reconstructed images can be readily available during data preparation, but to the best of our knowledge, they are surprisingly neglected in the past literature. We argue that they can provide extra information and build bridges for sparse- and full-view CT reconstruction.

To address the abovementioned issues, we propose a novel image-domain method for sparse-view CT reconstruction. 
\emph{First}, to address the limited receptive field inherent in conventional convolutional neural networks, we propose to learn \globalrepresent with fast Fourier convolution (FFC)~\cite{FFC}, so each element in \globalrepresent has an image-wide receptive field. 
This global nature allows artifacts and information, spread over the entire image, to be better modeled while also easing the alignment of the representations from different views. 
 \emph{Second}, to leverage extra supervision from intermediate-view CT images, we propose a novel distillation framework to learn better \globalrepresent,  termed \methodname, which contains a parallel teacher network to distill knowledge from the readily intermediate-view images to provide high-quality and appropriate guidance for learning the \globalrepresent of sparse-view CT images. Specifically, we first leverage intermediate-view reconstructed images to train a teacher network, which is then used to guide the learning of the student model (\ie the sparse-view CT).
The distillation scheme benefits \globalrepresent in two folds: (1) representation directional distillation that aligns the directions between the student and teacher \globalrepresent, which provides appropriate supervision considering the massive information loss due to the domain gaps between CT images with different views; 
and (2) band-pass-specific contrastive distillation that utilizes contrastive learning solely on the band-pass components to help distill the specific clinical value of each CT image without compromising the reconstruction accuracy.

\paragraph{Contributions.} In summary, our contributions are listed as follows.
\emph{First}, we propose the global representation (\globalrepresent) learning for sparse-view CT with Fourier convolution, which, to the best of our knowledge, is the first study to emphasize on the representation learning for image post-processing in sparse-view CT.
\emph{Second}, we propose a novel \globalrepresent distillation framework, which can leverage the extra supervision from intermediate-view reconstructed images for high-quality information recovery and reconstruction.
\emph{Third}, we present representation directional distillation and band-pass-specific contrastive distillation for distilling \globalrepresent to align the representation directions and gain clinically important details.
\emph{Last}, extensive experimental results demonstrate the superiority of \methodname over the state-of-the-art sparse-view CT reconstruction methods in terms of quantitative metrics and visual comparison.
%\end{enumerate}

%% file: section/related_work_detailed.tex
% !tex root=../main.tex

\section{Detailed Related Work}

\begin{figure*}[t] 
    \centering 
    \includegraphics[width=1.0\textwidth]{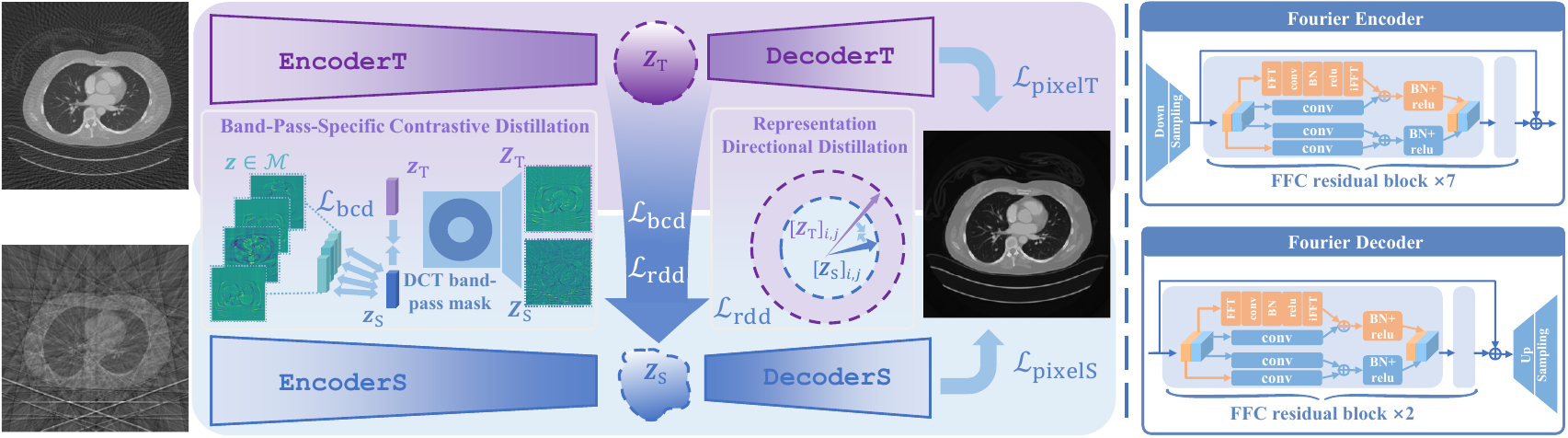}
    % \vspace{-10pt} 
    \caption{Overview of the proposed \methodname for sparse-view CT reconstruction. \methodname facilitates efficient information recovery by learning to distill the global representation from intermediate-view images.}
    \label{fig:flow} 
\end{figure*}

\subsection{Deep-learning-based Sparse-view CT}
Numerous model-based iterative reconstruction (IR) methods have been developed to solve the sparse-view CT problem, mainly utilizing the sparsity of the total variation (TV) of the image~\cite{ir_dbct, ir_piccs, ir_lowrank, ir_tv, ir_median, ir_admm, ir_l0}. However, most of them suffer from over-smoothed results, hand-crafted tuning for each image, and extremely high computation cost~\cite{dudotrans}, limiting their clinical application. 
In contrast, deep learning methods are often faster and more accessible, thus arousing attention in this field. Among them, most image-domain methods regard sparse-view CT reconstruction as an image post-processing task. 
RedCNN~\cite{redcnn}, FBPConvNet~\cite{fbpconvnet}, and DDNet~\cite{ddnet} are proposed to reconstruct the sparse-view CT by the convolutional neural network (CNN). These methods achieve competitive performance on streak artifact removal and structure preservation compared to conventional ones, but most of them fail to capture the global context and underperform in global reconstruction and artifact removal. 
Since streak artifacts result directly from the incomplete projection views, various methods are proposed to remove the artifacts by interpolating sinogram data~\cite{ssnet, sinogram_complete}. Further, dual-domain methods were becoming popular for their superior reconstruction performance by combining the knowledge of both domains. DuDoNet~\cite{dudonet} first proposed a novel Radon inversion layer linking the gradient between the image and sinogram domain networks. Furthermore, various techniques have been proposed to enhance dual domain approaches through network design~\cite{wnet, baimwcnn} and unrolling architecture~\cite{learn++}. Recently, Transformer~\cite{transformer} has been introduced to dual-domain methods for its capability of capturing long-range dependencies, achieving superior performance~\cite{dudotrans, ddptransformer, regformer}. However, the problems of the irreversible secondary artifacts and additional computational costs are not well addressed, and the requirement of raw data greatly limits their generalizability to other CT scanners/protocols. 
In this work, we challenge reconstructing sparse-view CT without raw data while still achieving state-of-the-art results.

\subsection{Knowledge Distillation}
Knowledge distillation transfers rich knowledge from teacher models to lightweight student models in the label or feature domains, mainly applied in model compression and acceleration~\cite{KD, KD2}.
Among various approaches, feature-based knowledge distillation is extensively applied in image restoration tasks such as image super-resolution~\cite{featuredis1,featuredis2,featuredis3,featuredis4}. It minimizes the distance between feature representations to learn richer information from the teacher model compared to softened labels~\cite{summary_dl}.
Further, contrastive representation distillation is proposed to exploit the structural characteristics among different samples by harnessing the discriminative representations, thereby enhancing differentiation within data~\cite{contrastdis, huang_dl, contrast_dl}.
This work distills knowledge from the intermediate-view images to enhance the sparse-view CT reconstruction.

\subsection{Frequency Methods in Deep Learning}
Frequency methods have been widely used in digital image processing and machine learning~\cite{digitalimageprocess}. 
Wavelet convolution~\cite{mwcnn} and fast Fourier convolution~\cite{FFC} have been proposed to provide a larger/global receptive field, achieving great success in image restoration tasks~\cite{lama}. Frequency methods have also yielded promising results in domain adaption and image translation, mainly by taking advantage of domain-invariant spectrum components~\cite{fda,fsdr,unsupda,imagetranslate,waveletstyle}. Previous research also shows that particular features and details can be better extracted in the frequency domain, leading to improvements in camouflaged object detection~\cite{camouflaged}, face forgery detection~\cite{forgery1,forgery12}, and face editing~\cite{wavletfaceedit}. 
Frequency-based networks have also made notable strides in CT reconstruction. Previous studies~\cite{leemwcnn, baimwcnn} employ multi-wavelet CNN~\cite{mwcnn} in both sinogram and image domain to suppress artifacts, substituting the pooling operator with wavelet transforms to achieve a larger receptive field. However, since both the Haar filter-based wavelet transform and convolution are applied on the spatial feature map, these methods struggle to capture global representation effectively. Recent research has demonstrated that employing convolution in the Fourier domain can efficiently eliminate the global artifacts~\cite{freeseed,quadnet}. 
In this work, we utilize a frequency-based network for global representation learning, enhancing clinically critical features by band-pass-specific contrastive distillation.

%% file: section/method.tex
% !tex root=../main.tex

\section{Method}

\subsection{Problem Definition}
Assume we have a two-dimensional (2D) CT slice of size $N\times N$, $\mat{I}\in\mathbb{R}^{N\times N}$, where each pixel contains the attenuation coefficient of the corresponding human body. The raw data or sinogram from the CT scanning, $\mat{S} \in \mathbb{R}^{N_\mathrm{v} \times N_\mathrm{d}}$, can be obtained via the Radon transform~\cite{radon_transform}, where $N_\mathrm{v}$ and $N_\mathrm{d}$ denote the number of projection views and the number of the detectors, respectively. Inversely, the image reconstruction process can be expressed as 
$
    \mat{I} = \mathcal{R}^{-1}(\mat{S})
$, where $\mathcal{R}^{-1}$ denotes the inverse Radon transform such as FBP. Ideally, when $N_\mathrm{v}$ is sufficiently large, FBP can produce pleasing image quality. However, when $N_\mathrm{v}$ is fairly small, the image reconstruction of so-called sparse-view CT becomes an undetermined problem. 

Here, we use $\mat{I}_{\mathrm{F}}$ and $\mat{I}_{\mathrm{S}}$ to represent the reconstructed images from full view (\ie ground-truth) and sparse view, respectively. The image-domain methods for sparse-view CT are to improve the image quality of $\mat{I}_{\mathrm{S}}$ towards $\mat{I}_{\mathrm{F}}$ through a neural network $\mathcal{F}$. That is, the output of the network, $\widehat{\mat{I}}_{\mathrm{S}} = \mathcal{F}(\mat{I}_{\mathrm{S}})$, is expected to be of comparable quality to the full-view ground-truth.
In addition, we also use the images, $\mat{I}_\mathrm{T}$,  reconstructed from an intermediate number of projection views that are between sparse and full views, as the teacher supervision, forming a student-teacher framework where the teacher encoder  with $\mat{I}_\mathrm{T}$  supervises the training of student encoder with $\mat{I}_\mathrm{S}$.

\subsection{Overview of Our \methodname}

Fig.~\ref{fig:flow} presents the proposed \methodname, which mainly consists of four parts: two encoders to learn global representations (\globalrepresent) from sparse- and intermediate-view reconstructed images, $\mat{I}_\mathrm{S}$, and $\mat{I}_\mathrm{T}$, respectively, and two decoders to produce the final processed images from  \globalrepresent. In addition to learning \globalrepresent with Fourier convolution, we have two novel modules to perform distillation from the teacher network:  
(1) representation directional distillation to align the directions between the student and teacher GloRes; and (2) band-pass-specific contrastive distillation to distill the clinically important features for better detail recovery.

Next, we detail each of the proposed components.

\subsection{Global Representation Learning}

The key technique behind the proposed network is the global representation learning with fast Fourier convolution (FFC)~\cite{FFC}. 
Concretely, FFC first applies real Fourier transform to get the frequency feature map and then performs convolution on frequency components before finally back transforming the frequency features as shown in Fig.~\ref{fig:flow}. Therefore, 
\globalrepresent from sparse-view images $\mat{I}_\mathrm{S}$ is learned by a Fourier-based encoder network as follows:
\begin{align}
\mat{Z}_\mathrm{S} = \opname{encoderS}(\mat{I}_\mathrm{S}),
\end{align}
where $\mat{Z}_\mathrm{S} \in \mathbb{R}^{N_\mathrm{w} \times N_\mathrm{h}\times N_\mathrm{c}}$ denotes \globalrepresent extracted from the sparse-view reconstructed image $\mat{I}_\mathrm{S}$ through the (student) encoder $\opname{encoderS}$ and $N_\mathrm{w} \times N_\mathrm{h}\times N_\mathrm{c}$ denotes the width, height, and channel size of \globalrepresent. 
Compared to the vanilla convolution with a limited receptive field, FFC works in the frequency domain, 
so each element in \globalrepresent contains the global information of the sparse-view images by using an image-wide receptive field, representing patterns of artifacts and the image contents simultaneously. 
Furthermore, such global nature can aid in modeling the artifact and information distributed throughout the image while easing the alignment between \globalrepresent of different views.

\subsection{Global Representation Distillation}

In the context of sparse-view CT, images with an intermediate view $\mat{I}_\mathrm{T}$ can be easily obtained from the full-view sinogram as extra supervision.
To get the teacher representation from $\mat{I}_\mathrm{T}$, a parallel teacher encoder $\opname{encoderT}$, using the same architecture as $\opname{encoderS}$ but with different weights, takes $\mat{I}_\mathrm{T}$ as input and yields teacher \globalrepresent $\mat{Z}_\mathrm{T}$ of the same shape with $\mat{Z}_\mathrm{S}$ as follows:
\begin{align}
\mat{Z}_\mathrm{T} = \opname{encoderT}(\mat{I}_\mathrm{T}).
\end{align}
$\mat{Z}_\mathrm{T}$ can be regarded as a high-quality pseudo target for student \globalrepresent in the same latent space since richer information can be extracted from $\mat{I}_\mathrm{T}$. Exploring distillation from $\mat{I}_\mathrm{T}$ is obviously easier than that of $\mat{I}_\mathrm{F}$ given the significant gap between $\mat{I}_\mathrm{S}$ and $\mat{I}_\mathrm{F}$.

Intuitively, $\mat{Z}_\mathrm{S}$ should be similar to $\mat{Z}_\mathrm{T}$ if the artifact-free target information is precisely encoded. Therefore, a natural way for distillation is to use $\mat{Z}_\mathrm{T}$ as the supervision for $\mat{Z}_\mathrm{S}$ by minimizing their Euclidean distance. However, this could harm the reconstruction performance since the student encoder has to partially recover $\mat{I}_\mathrm{T}$. \emph{Finding an effective solution to distilling the information in $\mat{Z}_\mathrm{T}$ for sparse-view CT reconstruction} is one of the highlights in this work.
Our idea is to separate the distillation process.
First, we present representation directional distillation to align the representation directions so the images of different views are along the same direction. Second, we propose a band-pass-specific contrastive distillation to enhance clinically critical details.

We detail these two distillation techniques as follows.

\subsubsection{Representation Directional Distillation}

Given that $\mat{I}_\mathrm{S}$ and $\mat{I}_\mathrm{T}$ share the same ground truth $\mat{I}_\mathrm{F}$, with the main difference being in the artifact patterns, student and teacher \globalrepresent should at least demonstrate a consistent direction for reconstruction with high confidence. 
Therefore, We define the directional distillation loss for $\mat{Z}_\mathrm{S}$ as follows:
\begin{align}
\mathcal{L}_\text{rdd} \!=\! \frac{1}{N_\mathrm{w}N_\mathrm{h}} \sum_{i,j} \Big(1 \!-\! \opname{CosSim}( [\mat{Z}_\mathrm{S}]_{i,j}, [\mat{Z}_\mathrm{T}]_{i,j})\Big), \label{eq:sup}% 
\end{align}
where $\opname{CosSim}$ denotes the cosine similarity, defined as $\opname{CosSim}(\vct{a},\vct{b})={\vct{a}\T\vct{b}}/{\max(\|\vct{a}\|_2 \|\vct{b}\|_2, \epsilon)}$ with $\epsilon$ set to $1\times 10^{-8}$ to avoid division by zero.

\subsubsection{Band-Pass-Specific Contrastive Distillation}

The directional distillation aligns the directions of student representation $\mat{Z}_\mathrm{S}$ and the teacher representation $\mat{Z}_\mathrm{T}$ in the view-independent space, which, however, does not guarantee a high-quality reconstruction of the image. 
What distinguishes sparse-view CT reconstruction task from classic image restoration one (\emph{e.g.} denoising, deraining, \emph{etc.}) is that the former is more demanding in terms of fine-grained details for diseases identification, organs delineation, and low-contrast lesions detection, \emph{etc}~\cite{clinicalselfsupCVPR}. 
In other words, it places more emphasis on learning the image-specific representation.

\begin{figure}[t] 
    \centering 
    \includegraphics[width=1\linewidth]{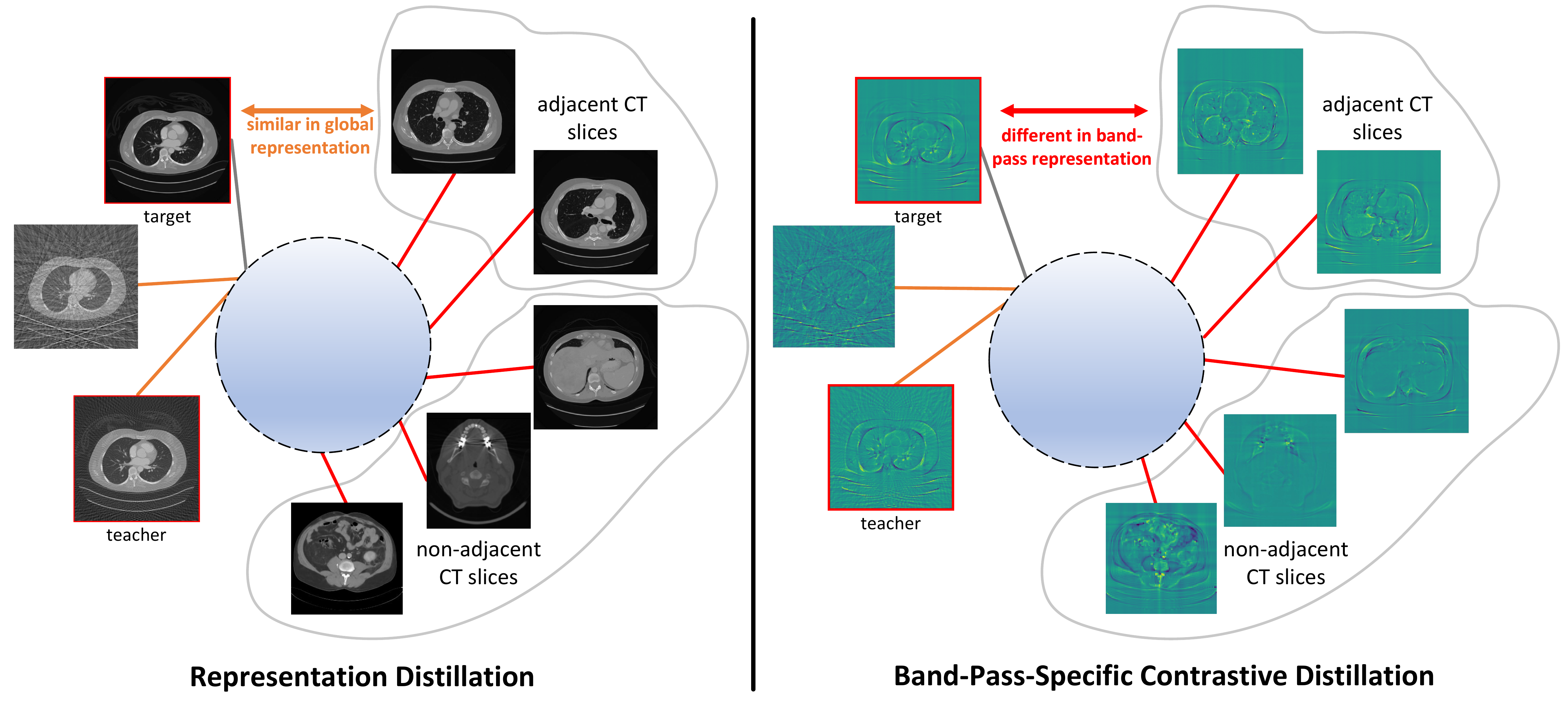}
    \caption{The difference between representation distillation (left) and band-pass-specific contrastive distillation (right). Global representation distillation fails to distinguish clinical details due to high similarity among similar body parts, \eg, adjacent lung slices. Band-pass-specific contrastive distillation further projects the representation to the DCT domain, enabling distillation of specific clinical details on corresponding DCT components.}
    \label{fig:bandscl} 
    % \vspace{-2mm}
\end{figure}

Contrastive representation learning, as a self-supervised approach, encourages the model to maximize the difference among representations of different samples so that model can emphasize those category-specific representations. However, directly implementing these losses would lead to imprecise reconstruction results because most CT images are inherently similar in intensity distribution, especially for images of the same body parts, as shown in Fig.~\ref{fig:bandscl}. 
Recent methods have achieved great success learning on specific DCT spectrum~\cite{fsdr,fda,dctdynamicsr}, due to its strong energy compaction property, making it simple to select specific frequency components~\cite{dct}.
Therefore, we distilled the image-specific details by applying contrastive distillation on specific discrete cosine transform (DCT) spectrums. 
On the one hand, detailed knowledge can be better distilled from specific frequency components of teacher \globalrepresent. On the other hand, contrastive against different CT slices can further enhance image-specific details on corresponding DCT components.

\paragraph{Discrete cosine transform.} 
Discrete cosine transform (DCT) is widely used for data compression in low-frequency components. The basis function of 2D DCT is defined as follows:
\begin{align}
B_{w,h}^{i,j} = \cos\left(\tfrac{\pi w}{N_\mathrm{w}}\left(i+\tfrac{1}{2}\right)\right)\cos\left(\tfrac{\pi h}{N_\mathrm{h}}\left(j+\tfrac{1}{2}\right)\right),
\end{align}
where $N_\mathrm{w}$ and $N_\mathrm{h}$ represent the  width and height of the input, respectively. Then, 2D DCT of the latent feature $\mat{Z}$ can be written as:
\begin{align}
f_{w,h}(\mat{Z})  =  \sum\nolimits_{i=0}^{N_\mathrm{w}-1} \sum\nolimits_{j=0}^{N_\mathrm{h}-1} [\mat{Z}]_{i,j} B_{w,h}^{i,j},
\end{align}
where $w \in \{0, 1,  \ldots, N_\mathrm{w}-1\}, h \in \{0, 1,\ldots, N_\mathrm{h}-1\}$.
The $f(\mat{Z}) \in \mathbb{R}^{N_\mathrm{w} \times N_\mathrm{h}}$ represents the 2D DCT frequency spectrum of $\mat{Z}$. Thus, frequency components can be easily selected by  a mask $\mat{M} \in \{0,1\}^{N_\mathrm{w} \times N_\mathrm{h}}$. 

\paragraph{Band-pass-specific contrastive distillation.}
Crucially, the low-frequency components of \globalrepresent contribute most to the reconstruction accuracy, while high-frequency may contain inherent noise and artifacts of CT images. We therefore use a band-pass supervised contrastive loss for the distillation, where a band-pass mask $\mat{M}$ is incorporated to filter out the mid-frequency part in the frequency domain, containing main structure as well as the necessary detail information. The proposed band-pass supervised contrastive loss can be written as:
\begin{align}
\vct{z} = \opname{flatten}\left(\opname{crop}\left(f_{w,h}(\mat{Z}) ,\mat{M}\right)\right),
\end{align}
where $\opname{crop}$ operator will keep the value at locations where $\mat{M}=1$ while the $\opname{flatten}$ operator flattens the 2D spectrum map into 1D representation. 

It is to be observed that the clinical details in the CT images vary from case by case and are each very critical for diagnosis, requiring more sophisticated instance-level distinction. To this end, we consider all representations from different CT slices as negative samples to maximize image-specific features while updating the memory bank $\mathcal{M}$ with the widely applied First-In-First-Out mechanism per iteration from the previous mini-batches. 
The proposed band-pass-specific contrastive distillation loss for sample $\vct{z}_\mathrm{S}$ can be written as follows:
\begin{align}
    \mathcal{L}_{\mathrm{bcd}}  =-\log{\frac{\text{exp}\left(\vct{z}_\mathrm{S}\T\vct{z}_\mathrm{T}/\tau\right)}{\sum\nolimits_{\vct{z}\in \mathcal{M}}\text{exp}\left(\vct{z}_\mathrm{S}\T\vct{z}/\tau\right)}},\label{eq:bpSCD}
\end{align}
where $\vct{z}_\mathrm{S}$, $\vct{z}_\mathrm{T}$, and $\vct{z}$ are frequency representations of $\mat{Z}_\mathrm{S}$, $\mat{Z}_\mathrm{T}$, and negative samples from memory bank $\mathcal{M}$ of size $N_\mathrm{mem}$, respectively. $\tau$ is the temperature term used to adjust the sensitivity of negative samples.

\subsection{Loss Function and Training Procedure}
\paragraph{Teacher reconstruction.}
To constrain the \globalrepresent extracted from $\mat{I}_\mathrm{S}$ and $\mat{I}_\mathrm{T}$ in the same latent space, a naive solution is to share the same decoder in the training phase, which, however, brings instability. Therefore, we adopt the widely used Exponential Moving Average (EMA) with momentum $m$ to update the parameters $\theta$ of the teacher decoder by the student decoder~\cite{mocov1}:
\begin{align}
\theta_{\opname{decoderT}} = m\theta_{\opname{decoderT}} + (1-m)\theta_{\opname{decoderS}}. \label{eq:EMA} 
\end{align}
Then, the global representation $\mat{Z}_\mathrm{T}$ learned by the teacher is fed into to the \decoder to obtain the teacher reconstruction result $\widehat{\mat{I}}_{\mathrm{T}}$, and we use $\ell_1$ loss to measure the pixel-wise difference between $\widehat{\mat{I}}_{\mathrm{T}}$ and the full-view ground-truth $\mat{I}_{\mathrm{F}}$ and train the teacher:
\begin{align}
\mathcal{L}_\mathrm{pixelT} = \|\opname{decoderT}({\mat{Z}}_\mathrm{T}) - {\mat{I}}_\mathrm{F} \|_1.\label{eq:reconT}
\end{align}

\paragraph{Student reconstruction.} 
The pixel-wise error measurement for the  student is defined in a similar way:
\begin{align}
\mathcal{L}_\mathrm{pixelS} = \|\opname{decoderS}(\mat{Z}_\mathrm{S}) - \mat{I}_\mathrm{F} \|_1.\label{eq:reconS}
\end{align}
Finally, a compound loss function combining pixel-wise error and two distillation losses respectively defined in Eq.~\eqref{eq:sup} and Eq.~\eqref{eq:bpSCD} is introduced for student network training to achieve high-quality sparse-view CT reconstruction: 
\begin{align}
    \mathcal{L}_{\mathrm{Stu}} = \mathcal{L_{\mathrm{pixelS}}} + \alpha \mathcal{L}_{\mathrm{rdd}} + \beta \mathcal{L}_{\mathrm{bcd}},\label{eq:stu}
\end{align}
\methodname is trained following the process summarized in Algorithm~\ref{algorithm}, where the student and \teacherencoder are trained iteratively.

\begin{algorithm}[t]
    \renewcommand{\algorithmicensure}{\textbf{Return:}}
    \caption{The training process of \methodname.}
    \label{algorithm}
    \begin{algorithmic}[1]
        \REQUIRE  $\opname{encoderS}$, $\opname{encoderT}$, $\opname{decoderS}$, $\opname{decoderT}$;
        Input images: $\mat{I}_\mathrm{S}$, $\mat{I}_\mathrm{T}$, $\mat{I}_\mathrm{F}$; \\
        momentum $m$ and Memory bank
        \FOR{$iter = 0$ \textbf{to} $Max\_Iter$}
        \STATE Update $\opname{decoderT}$ using Eq.~\eqref{eq:EMA}
        \STATE $\widehat{\mat{I}}_\mathrm{T}, \mat{Z}_\mathrm{T}= \opname{decoderT}(\opname{encoderT}(\mat{I}_\mathrm{T}))$
        \STATE Optimize $\opname{encoderT}$ using Eq.~\eqref{eq:reconT}
        \STATE $\widehat{\mat{I}}_\mathrm{S}, \mat{Z}_\mathrm{S}= \opname{decoderS}(\opname{encoderS}(\mat{I}_\mathrm{S}))$
        
        \STATE Optimize $\opname{encoderS}$+$\opname{decoderS}$ using Eq.~\eqref{eq:stu} 
        \STATE Update memory bank
        \ENDFOR \\
    \ENSURE $\opname{encoderS}$+ $\opname{decoderS}$
    \end{algorithmic}
    \end{algorithm}

%% file: section/result.tex
% !tex root=../main.tex

\section{Experimental Results}

\begin{table*}
\centering
\small
\begin{tabular}{l|rrr|rrr|rrr|rrr}
&
\multicolumn{3}{c|}{$N_\mathrm{v}$ = 18}& 
\multicolumn{3}{c|}{$N_\mathrm{v}$ = 36}& 
\multicolumn{3}{c|}{$N_\mathrm{v}$ = 72}& 
\multicolumn{3}{c}{$N_\mathrm{v}$ = 144}\\ 
Methods & PSNR & \multicolumn{1}{l}{SSIM} & RMSE 
& \multicolumn{1}{l}{PSNR} & \multicolumn{1}{l}{SSIM} & RMSE 
& \multicolumn{1}{l}{PSNR} & SSIM & RMSE & PSNR & SSIM & RMSE\\ 
\shline
FBP       & 22.22& 35.36& 0.0795 & 25.49& 47.49& 0.0543& 30.88& 63.81 & 0.0293& 37.12 & 82.94 & 0.0144 \\
DDNet~\cite{ddnet} & 34.57& 91.94& 0.0187& 38.24& 94.91& 0.0126& 41.66& 97.27 & 0.0083& 46.75& 98.79 & 0.0048 \\
FBPConvNet~\cite{fbpconvnet}     & 35.95& 93.62& 0.0164& 39.79& 96.13 & 0.0106& 43.76 & 97.47 & 0.0066  & 48.46 & 98.60 & 0.0039 \\
DuDoNet~\cite{dudonet}      & 35.69& 93.96 & 0.0169 & 40.36& 96.94& 0.0103  & 44.86 & 98.39 & 0.0059 & 49.33& 99.23 & 0.0036 \\
DDPTrans~\cite{ddptransformer}          & 35.11 & 93.48  & 0.0181& 38.68  & 95.99  & 0.0121& 43.56 & 98.16 & 0.0069 & 48.72 & 99.22 & 0.0038 \\
DuDoTrans~\cite{dudotrans}  & 36.08 & 93.28  & 0.0161 & 40.75 & 96.67& 0.0095 & 45.16  & \textbf{98.44} & \textbf{0.0057}  & \textbf{49.96}  & \textbf{99.28} & \textbf{0.0034} \\
\hline
\backbonename (ours)            & 38.08 & 95.20 & 0.0129 & 40.86 & 96.81 & 0.0093 & 44.43 & 98.18 & 0.0063 & 48.45 & 99.08 & 0.0039 \\
\methodname (ours) & \textbf{38.65} & \textbf{95.87} & \textbf{0.0120} & \textbf{41.25} & \textbf{97.05} & \textbf{0.0090} & \textbf{45.18} & 98.43 & \textbf{0.0057} & 48.96 & 99.21 & 0.0037 \\       
\end{tabular}
\caption{Quantitative evaluation [PSNR (db), SSIM (\%) and RMSE] for state-of-the-art methods on DeepLesion dataset.}
\label{tab:state_of_the-art}
\end{table*}

\begin{figure*}[!htb]
    \centering
    \includegraphics[width=1\linewidth]{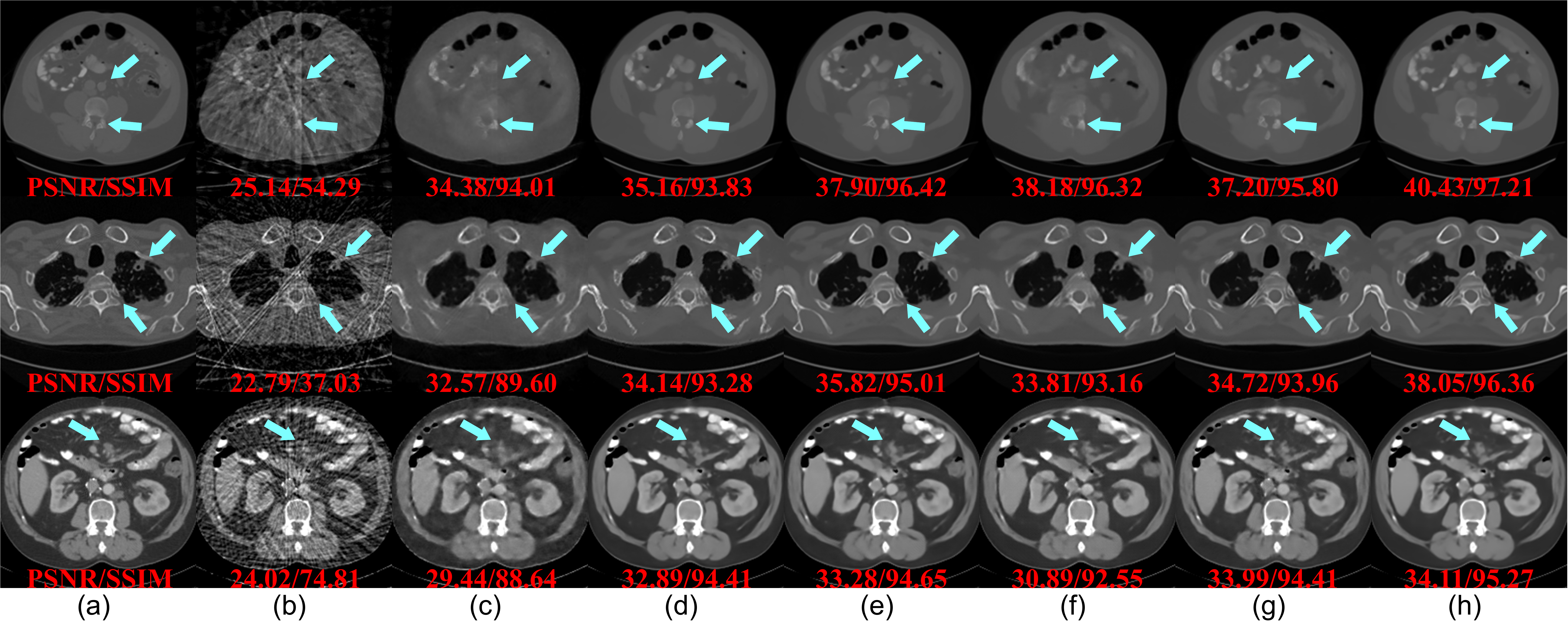}
    \vspace{-12pt}
    \caption{Visual comparison of state-of-the-art methods on DeepLesion dataset: (a) Ground Truth, (b) FBP, (c) DDNet, (d) FBPConvNet, (e) DuDoNet, (f) DDPTrans, (g) DuDoTrans, and (h) \methodname. From top to bottom:  the results under $N_\mathrm{v}=18, 36, 72$;  display window is set to [-1000, 2000] HU for the first two rows and [-200, 300] HU for the third row.}
    \label{fig:deep} 
\end{figure*}

\subsection{Experimental Setup}
\paragraph{Dataset.}
We use DeepLesion dataset~\cite{deeplesion}  and ``2016 NIH-AAPM-Mayo Clinic Low-Dose CT Grand Challenge''  AAPM dataset~\cite{aapm} to demonstrate the effectiveness of the proposed \methodname.
The DeepLesion dataset is the largest multi-lesion real-world CT dataset made available to the public, from which 40,000 images of 303 patients are selected as the training set while 1000 images of another 18 patients are selected as the test set.
AAPM dataset contains routine dose CT data from 10 patients, where a total of 5,410 slices from 9 patients are chosen for training and 526 slices from the remaining 1 patient for testing. All images are resized to $256\times 256$. We simulate the forward and back projection using fan-beam geometry under 120 kVp and 500 mA with TorchRadon toolbox~\cite{torch_radon}. The distance from X-ray source to the rotation center is 59.5cm, and the number of detectors is set to 672. Sparse-view CT images are generated from $N_\mathrm{v} = 18, 36, 72, 144$ projection views uniformly sampled from full 720 views covering $[0,2\pi]$. To simulate the photon noise presented in real-world CT, an intensity of $1\times10^6$ Poisson noise is added to the sinograms. 

\paragraph{Implementation details.}
Our model is implemented in PyTorch. 
We use Adam optimizer~\cite{adam} with $(\beta_1,\beta_2)=(0.5,0.999)$ to train the model. The learning rate starts from $1\times 10^{-3}$ and is halved for every 40 epochs. For each method, four models of different $N_\mathrm{v}$ are trained separately on 4 NVIDIA RTX 3090 GPUs for 120 epochs with a batch size of 8. 
For hyperparameters in \methodname, we set $\alpha$ and $\beta$ in Eq.~\eqref{eq:stu} empirically to 0.1 and 0.0002 to fit the scale between losses. The momentum $m$ and temperature $\tau$ are set to 0.9 and 1.0, respectively, while the size $L$ of the memory bank is set to 300 to balance the performance and computational cost. We set intermediate-view as $N_\mathrm{v} \times 2$ by default. More details, including the network architecture, can be found in the Appendix. 

\paragraph{Evaluation metrics.}
For quantitative evaluation, we use peak signal-to-noise ratio (PSNR), structural similarity (SSIM)~\cite{ssim} and root mean square error (RMSE); all of them are widely adopted for image quality assessment.

\begin{table*}[htbp]
\centering
\small
\begin{tabular}{l|rrr|rrr|rrr|rrr}
&
\multicolumn{3}{c|}{$N_\mathrm{v}$ = 18}&
\multicolumn{3}{c|}{$N_\mathrm{v}$ = 36}&
\multicolumn{3}{c|}{$N_\mathrm{v}$ = 72}&
\multicolumn{3}{c}{$N_\mathrm{v}$ = 144} \\ 
%\cline{2-13}
Methods& PSNR & SSIM & RMSE & 
PSNR & {SSIM} & RMSE 
& {PSNR} & SSIM & RMSE & PSNR & SSIM & RMSE\\ 
\shline
FBP            & 22.73 & 35.06 & 0.0732 & 26.27 & 47.52 & 0.0486 & 31.36 & 64.56 & 0.0270 & 37.85 & 84.57 & 0.0128 \\
DDNet~\cite{ddnet}          & 34.29 & 89.68 & 0.0194 & 36.95 & 93.03 & 0.0142 & 40.41 & 96.04 & 0.0096 & 44.35 & 98.05 & 0.0061 \\
FBPConvNet~\cite{fbpconvnet}     & 35.73 & 92.84 & 0.0165 & 37.95 & 93.73 & 0.0127 & 42.92 & 97.19 & 0.0072 & 47.35 & 98.76 & 0.0043 \\
DuDoNet~\cite{dudonet}        & 34.82 & 93.00 & 0.0196 & 39.89 & 95.99 & 0.0102 & 44.06 & 98.02 & \textbf{0.0062} & 48.39 & 99.11 & 0.0038 \\
DDPTrans~\cite{ddptransformer} & 34.47 & 91.95 & 0.0191 & 38.13 & 94.83 & 0.0125 & 42.77 & 97.65 & 0.0073 & 47.84 & 99.06 & 0.0041 \\ 
DuDoTrans~\cite{dudotrans}      & 35.85 & 93.07 & 0.0162 & 40.04 & 96.02 & 0.0100 & 44.20 & \textbf{98.07} & \textbf{0.0062} & \textbf{49.02} & \textbf{99.21} & \textbf{0.0036} \\
\hline
\backbonename (ours)  & 37.20 & 93.90 & 0.0140 & 41.46 & \textbf{96.84} & 0.0085 & 43.64 & 97.84 & 0.0066 & 47.50 & 99.03 & 0.0042 \\
\methodname(ours) & \textbf{37.91} & \textbf{94.58} & \textbf{0.0128} & \textbf{41.57} & 96.70 & \textbf{0.0084} & \textbf{44.24} & 98.03 & \textbf{0.0062} & 48.10 & 99.08 & 0.0040  
\end{tabular}
\caption{Quantitative evaluation [PSNR (db), SSIM (\%) and RMSE] for state-of-the-art methods transferred to AAPM dataset.}
\label{tab:sota-aapm}
\vspace{-2mm}
\end{table*}

\begin{figure*}[!htb]
    \centering
    \includegraphics[width=1\linewidth]{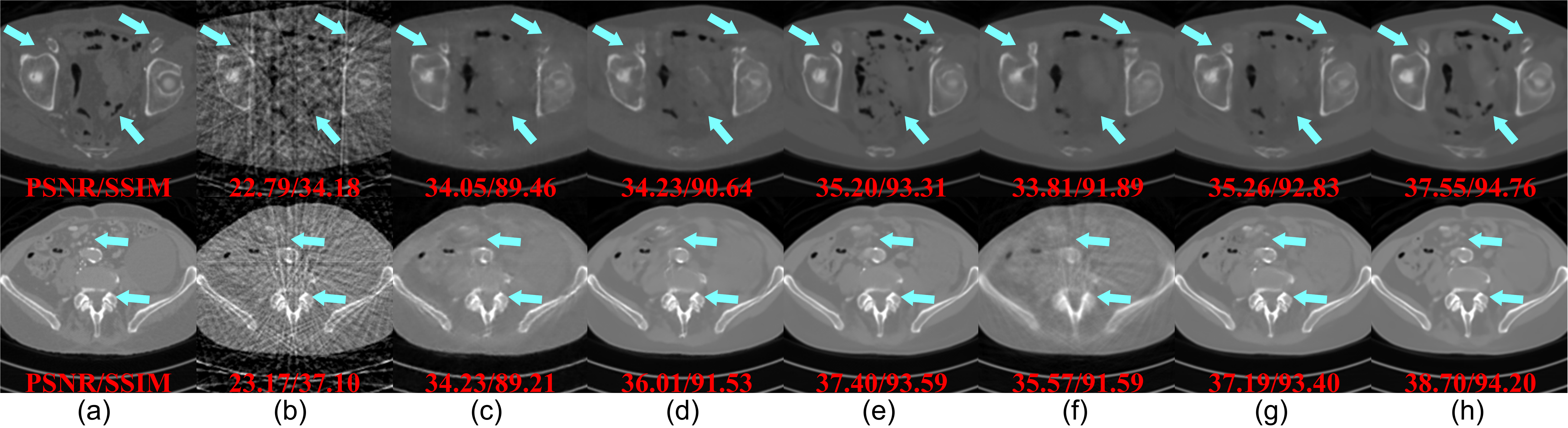} 
    \vspace{-12pt}
    \caption{Visual comparison of state-of-the-art methods on AAPM dataset: (a) Ground Truth, (b) FBP, (c) DDNet, (d) FBPConvNet, (e) DuDoNet, (f) DDPTrans, (g) DuDoTrans, and (h) \methodname. From top to bottom: $N_\mathrm{v}=18$, $36$;
    the display window is [-1000, 2000]~HU. } 
    \label{fig:aapm} 
    \vspace{-2mm}
\end{figure*}

\subsection{Comparison with State-of-the-Art Methods}
We compare \methodname with the following state-of-the-art methods: {DDNet}~\cite{ddnet}, {FBPConvNet}~\cite{fbpconvnet}, {DuDoNet}~\cite{dudonet}, {DDPTrans}~\cite{ddptransformer}, and {DuDoTrans}~\cite{dudotrans}. In addition, we name the network without the distillation from $\mat{I}_\mathrm{T}$ as the frequency encoder and decoder network (\backbonename), optimized with only the pixel-wise loss in Eq.~\eqref{eq:reconS}.
FBP directly applies Radon transform to reconstruct the image from the sparse-view sinogram. DDNet and FBPConvNet are image-domain deep-learning methods taking FBP results as input. DuDoNet is a state-of-the-art dual-domain method that restores the CT image using two U-Nets in sinogram and image domains. Furthermore, DDPTrans and DuDoTrans adopt Transformer on both domains to restore the image by utilizing the long-range dependency. 
We try our best to reproduce these methods and then train and test them on the same dataset for a fair comparison.

\paragraph{Quantitative comparison.}
Table~\ref{tab:state_of_the-art} presents the quantitative comparison. Generally, dual-domain methods perform better than single domain methods, especially in the case of $N_\mathrm{v}=72, 144$ thanks to the sinogram inpainting network. 
However, we highlight in ultra sparse-view scenarios where $N_\mathrm{v}= 18$, that dual-domain methods like DuDoNet perform worse than classical image-domain-only methods such as FBPConvNet. 
This is mainly due to the secondary artifact introduced by the unsuccessful sinogram inpainting; \eg, when $N_\mathrm{v}=18$ in our setting, directly restoring the sinogram can be recognized as a 40$\times$ super-resolution task, which is rather challenging. 
Interestingly, DuDoTrans is superior to DuDoNet to some extent, indicating the importance of long-range dependency in sparse-view CT reconstruction. DDPTrans fails to surpass DuDoNet mainly because of the limited parameter number though most transformers are memory-consuming.

Compared with the previous works, the proposed \methodname achieves competitive results by learning to distill \globalrepresent for high-quality reconstruction.
Particularly in ultra-sparse scenarios, \eg, when $N_\mathrm{v}=18$, \methodname has 2.57db and 2.70db improvement on PSNR over DuDoTrans and FBPConvNet, respectively. As $N_\mathrm{v}$ increases, the advantage of \methodname decreases, mainly due to a shift towards artifacts removal tasks rather than reconstruction.
We highlight that the fewer projection views indicate less radiation dose and more speedup. 
We notice that global representation distillation further improves the performance of \methodname from 0.51db to 0.89db compared to \backbonename. 
The results show that the proposed \methodname successfully inspire the potential of image post-processing methods and is significantly superior to the state-of-the-art methods under various settings.

\paragraph{Visual comparison.}
Fig.~\ref{fig:deep} presents the visualization results of three sparse-view images. DDNet and FBPConvNet cannot remove the artifact thoroughly, while the reconstruction results are over-smooth in the boundary of organs and bones. Such effects can be easily observed in the second row of Figs.~\ref{fig:deep}\textbf{(c)} and~\ref{fig:deep}\textbf{(d)}  on the spine where the anatomical structure is complicated. Although dual-domain methods perform better in general, we find that the accuracy in details is even worse than FBPConvNet in comparison with Figs.~\ref{fig:deep}\textbf{(d)} and~\ref{fig:deep}\textbf{(e)}-\textbf{(g)}. 

Among all these methods, \methodname best recovers the structures and details, especially in ultra-sparse scenarios. As shown in the first two rows of Fig.~\ref{fig:deep}\textbf{(h)}, only \methodname precisely reconstructs those clinically important details such as soft tissue and the corrupted pulmonary alveoli.

\subsection{Ablation Study}

We first evaluate the effectiveness of each component in \methodname. 
We use \backbonename as the baseline to add or remove components. 
The configurations involved in Table~\ref{tab:ablation} are mainly four groups: (1) the baseline \backbonename using pixel-wise loss to train student model without distillation components (\textbf{a}); (2) the baseline without Fourier convolutions (\textbf{b}--\textbf{c}); (3) the baseline with distillation from intermediate-view reconstructed images (\textbf{d}--\textbf{g}); and (4) the dual-domain version with raw data processed by sinogram-domain sub-network of DuDoNet~\cite{dudonet} (\textbf{h}).
Unless noted otherwise, pixel-wise loss is involved in training.

\begin{table}
    \centering
    \small
    \begin{tabular}{r|l|c}
    & config.  & PSNR \\
    \shline
    \textbf{a}) & \backbonename             & 38.09 \\ \hline
    \textbf{b}) & \backbonename w/o Fourier & 36.52 \\
    \textbf{c}) & \backbonename w/o Fourier + $\mathcal{L}_\mathrm{rdd}$ + $\mathcal{L}_{\mathrm{bcd}}$& 36.30 \\ \hline
    \textbf{d}) & \backbonename + \text{data augmentation} & 34.41 \\
    \textbf{e}) & \backbonename + $\mathcal{L}_\mathrm{rdd}$            & 38.39 \\
    \textbf{f}) & \backbonename + $\mathcal{L}_{\mathrm{bcd}}$ & 38.42 \\
    \textbf{g}) & \backbonename + $\mathcal{L}_\mathrm{rdd}$ + $\mathcal{L}_{\mathrm{bcd}}$ =  \textbf{\methodname} & \textbf{38.65} \\
    \hline
    \textbf{h}) & \backbonename + \text{raw data} & 38.44 \\
    \end{tabular}
    \caption{Quantitative evaluation of different configurations.}
    \label{tab:ablation}
    \vspace{-2mm}
\end{table}

\paragraph{Ablations on global representations.}
The results between \textbf{a)} and \textbf{b)} in Table~\ref{tab:ablation} confirm that learning global representations with the image-wide receptive field by Fourier convolution is beneficial for sparse CT reconstruction.
Also, we notice that in Tables~\ref{tab:state_of_the-art} and~\ref{tab:sota-aapm}, our baseline \backbonename still outperforms state-of-the-art methods when $N_\mathrm{v}=18$ and  $36$.
Interestingly, we found that \textbf{c)} performs even worse than \textbf{b)} in Table~\ref{tab:ablation}. This suggests that the global representation is essential for bridging the gap between images of different views. Generally, learning global representations is powerful and provides a new perspective for sparse-view CT reconstruction, which also answers that the limit of image post-processing methods is still far beyond reach.

\paragraph{Ablations on  global representation distillation.}
Through a comparison between \backbonename  and the one trained with intermediate-view images as data augmentation (\textbf{a} vs.~\textbf{d}) in Table~\ref{tab:ablation}, we found that directly applying data augmentation had a negative impact, resulting in a 3.67db PSNR drop. This indicates that the network failed to benefit from denser-view images directly because of the significant domain gap among images of different view. By comparing \textbf{a)}, \textbf{e)} and \textbf{f)} in Table~\ref{tab:ablation}, we can see that both \textbf{e)} representation directional distillation and \textbf{f)} band-pass-specific contrastive distillation can improve the performance of \textbf{a)}, showing that end-to-end supervision cannot thoroughly unlock the potential of the \globalrepresent. 
By training with both distillation loss, \textbf{g)} further boosts the performance and outperforms all other configurations. By comparing \textbf{a)}, \textbf{g)} and \textbf{h)}, we found unsurprisingly that using raw data does improve the performance of the image-domain network. However, our method \methodname yields even better performance by effectively distilling \globalrepresent from intermediate-view image data.

\paragraph{Ablations on intermediate views for distillation.}
Selecting a suitable intermediate view for distillation is crucial in realizing the full potential of \methodname. Images reconstructed from a denser view can provide richer information to the teacher \globalrepresent while introducing a more significant domain gap.
As presented in Table~\ref{tab:ablation_views}, we found that models distilled from $N_\mathrm{v} \times 2$ views exhibit the best performance. Interestingly, models distilled from the full-view image yield the worst performance, primarily because of the considerable domain gap between the input data.

Further ablation studies for \emph{framework designs}, \emph{configurations of residual blocks}, and \emph{ablation study for distillation loss} can be found in the Appendix.

\begin{table}
    \centering
    \small
    \begin{tabular}{c|ccccc}
        config.
        & $N_\mathrm{v} \times 2$  & $N_\mathrm{v} \times 3$ 
        & $N_\mathrm{v} \times 4$ & 720 \\
    \shline
    $N_\mathrm{v}=18$ & \textbf{38.38} & 38.29 & 38.28 & 38.02 \\
    $N_\mathrm{v}=72$ & \textbf{44.84} & 44.46 & 44.72 & 44.38 \\
    \end{tabular}
    \caption{PSNR evaluation of \methodname trained with different intermediate-view images for 60 epochs. The first and second rows show the results of sparse-view images with $N_\mathrm{v}=18, 72$ distilled from images of $N_\mathrm{v} \times 2, 3, 4$ and $720$ (full view), respectively.}
    \label{tab:ablation_views}
    \vspace{-2mm}
\end{table}

\subsection{Transfer to Other Dataset}
To test the generalizability and robustness of \methodname along with other state-of-the-art methods, we finetune each model for another ten epochs on AAPM dataset to bridge the domain gap following the same setting.
In Table~\ref{tab:sota-aapm}, the proposed \methodname shows excellent transferability over all other state-of-the-art methods in (ultra) sparse-view scenarios ($N_\mathrm{v}=18,36,72$) and still achieves performance comparable with the dual-domain methods when $N_\mathrm{v}=72$. Specifically, when $N_\mathrm{v}=18$, we notice \methodname has 2.06db and 2.18db improvements on PSNR compared to DuDotrans and FBPConvNet, respectively, indicating that our method is significantly superior in ultra sparse-view scenarios. 
Fig.~\ref{fig:aapm} shows images with $N_\mathrm{v}=18$ and $36$, which are corrupted severely by FBP as shown in Fig.~\ref{fig:aapm}\textbf{(b)}, thus losing its clinical value. Previous image-domain methods remove the artifact to some extent but lose most of the details, as shown in Figs.~\ref{fig:aapm}\textbf{(c)} and \textbf{(d)}. Although the state-of-the-art dual-domain methods gain better results in general, we notice that clinical details such as lesion and bone, as pointed out by the blue arrows, are not accurately reconstructed in  Figs.~\ref{fig:aapm}\textbf{(e)}-\textbf{(g)} while severe secondary artifacts are introduced, especially when $N_\mathrm{v}=18$. In contrast, the proposed \methodname excels at reconstructing the details and greatly improves the clinical value of sparse-view CT.

%% file: section/conclusion.tex
% !tex root=../main.tex

% \section{Conclusion}
\section{Discussion and Conclusion}

First, we emphasize that our contributions can be easily extended to dual-domain methods when sinogram data are available. On the one hand, learning to distill \globalrepresent can also benefit the image-domain network in a dual-domain framework. On the other hand, the proposed method can also be used to enhance the sinogram recovery by distilling the sinogram with intermediate views. Future work can concentrate on extending \methodname to dual domain methods and other CT reconstruction tasks, such as limited-views CT reconstruction and metal artifact reduction. 
Second, although the teacher network is trained alongside the student network in the current version, it is valuable to explore the utilization of pre-trained teacher models, enabling \methodname to progressively distill the \globalrepresent from a series of teacher models.

We also acknowledge some limitations. First, the teacher network will introduce additional computational costs during the training phase. It is possible to study how to distill knowledge from a pretrained teacher network.
Second, although there are clear improvements in quantitative and visual results, it is desirable to have feedback from radiologists for clinical practice. The generalizability and robustness of the trained networks in clinical applications require further investigation.  
Third, we only considered 2D cases in the paper due to memory respect. If GPU memory permits, one can extend \methodname to 3D by replacing 2D convolutions with 3D ones by applying high-dimension kernels for FFCs~\cite{FFC}. 
Lastly, \methodname failed to surpass the dual-domain methods in denser view scenarios when sinogram can provide effective information for performance improvement. Studying how to improve the performance given an arbitrary $N_\mathrm{v}$ would be of great value for future research.

% conclusion
This paper sticks to reconstructing the sparse-view CT directly in the image domain by learning to distill global representation from intermediate-view reconstructed images innovatively. 
We propose to learn \globalrepresent with Fourier convolution for image-wide receptive field and present directional and band-pass-specific contrastive distillation for high-quality reconstruction.
The proposed framework \methodname achieves state-of-the-art performance without access to sinogram data, demonstrating that the limit of the image post-processing methods is still far beyond reach.

%% file: Supplementary.tex
\appendix

\renewcommand \thepart{}
\renewcommand \partname{}
\renewcommand{\thetable}{S\arabic{table}}
\renewcommand{\thefigure}{S\arabic{figure}}
\addcontentsline{toc}{section}{Appendix}
\numberwithin{equation}{section}
\setcounter{figure}{0}
\setcounter{table}{0}
\renewcommand\thetable{A\arabic{table}}
\renewcommand\thefigure{A\arabic{figure}}

\crefname{section}{Sec.}{Secs.}
\Crefname{section}{Section}{Sections}
\Crefname{table}{Table}{Tables}
\crefname{table}{Tab.}{Tabs.}

\clearpage

\part{\hfill \textsc{Appendix} \hfill}

% \begin{abstract}
This Appendix includes five parts: 
(A) more ablation study and analysis,
(B) efficiency,
(C) more visualization results, 
and (D) detailed network architectures.
% \end{abstract}

\input{./section/more_ablation.tex}

\input{./section/more_details.tex}

%% file: section/more_ablation.tex
% !tex root=../Supplementary.tex

\section{More Ablation Study and Analysis}

\subsection{Ablation on Framework Design}
\begin{figure}[htb] 
    \centering 
    \includegraphics[width=0.45\textwidth]{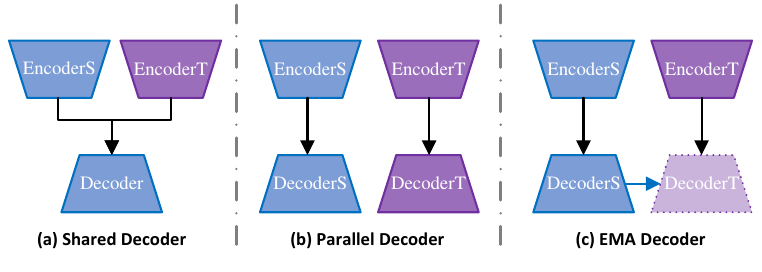}
    \caption{Framework designs of \methodname: (a) \methodname-\textsc{S}: shared decoder for student and teacher networks, which is optimized twice per iteration; (b) \methodname-\textsc{P}: separate parallel decoders for student and teacher; and (c) \methodname-\textsc{E}: teacher decoder is updated via EMA according to student decoder. For simplicity, we name \methodname-\textsc{E} trained without distillation loss as \methodname-\textsc{N} for comparison.} 
    \label{fig:ablation_framework} 
    \vspace{-2mm}
\end{figure}
Table~\ref{tab:ablation_framework} presents the experimental results of different framework designs employed in \methodname, as illustrated in Fig.~\ref{fig:ablation_framework}.
We found that the parallel decoder design failed to align the representations of different views into a shared latent space, thus, leading to suboptimal results. The result of \methodname-\textsc{P} is even worse than \methodname-\textsc{N}, revealing that the domain gap between various views can harm the training.
Although the utilization of a shared decoder can enforce the representation to be in the same latent space and outperform \methodname-\textsc{P}, it could also introduce unstable problems during training, for the parameters of the shared decoder are updated twice in an iteration.
By contrast, \methodname-\textsc{E} achieves the alignment of representations in a shared latent space through an exponential moving average (EMA) update procedure, thereby circumventing interference with student training. The resultant teacher encoder can be considered a stable version of the student encoder, making the teacher \globalrepresent a dependable distillation target. Therefore, we choose \methodname-\textsc{E} as our final framework design.
\begin{table}[htb]
    \centering
    \small
    \begin{tabular}{c|ccccc}
    \methodname  & -\textsc{N} & -\textsc{P}  & -\textsc{S} & -\textsc{E}  \\
    \shline
    PSNR  & 37.91  & 37.85 & 38.04 & \textbf{38.38} \\
    \end{tabular}
    \caption{PSNR evaluation of \methodname with different framework designs. All networks are trained under $N_\mathrm{v}=18$ for 60 epochs.}
    \label{tab:ablation_framework}
\end{table}

\subsection{Ablation on Configurations of Residual Blocks}
\begin{table}[htb]
    \centering
    \small
    \begin{tabular}{c|ccccc}
    config.  & e5d4 & e6d3  & e7d2  & e8d1  \\
    \shline
    $N_\mathrm{v}=18$ & 37.49 & 37.62  & \textbf{38.06} & 38.02  \\
    $N_\mathrm{v}=72$ & 43.75 & 43.90  & \textbf{44.39} & 44.08  \\
    \end{tabular}
    \caption{PSNR evaluation of \methodname with varied numbers of FFC residual blocks in the encoder and decoder. (\eg, e5d4 represents 5 and 4 FFC residual blocks in the encoder and decoder, respectively). All models are trained for 40 epochs considering the computational cost.}
    \label{tab:ablation_blocks}
\end{table}
Given the fixed parameters, a larger encoder can improve the information extraction and recovery process, as well as better bridge the domain gap between the sparse- and denser-view images. In the meantime, a larger decoder can better decode the global representation and improve the reconstruction quality. 
Table~\ref{tab:ablation_blocks} presents the quantitative results of varying numbers of FFC residual blocks in the encoder and decoder. The results suggest that a ratio of $7:2$ for 9 residual blocks in the encoder and decoder is the most favorable for distillation.

\subsection{Ablation on Distillation Loss}
\begin{table}[htb]
    \centering
    \small
    \setlength{\arraycolsep}{2pt}
    \begin{tabular}{c|ccc}
    config.  & $\ell_1$ loss & $\ell_2$ loss & (ours) \\
    \shline
    $N_\mathrm{v}=18$ & 37.44 & 37.29  & \textbf{38.06}  \\ % nothing 37.80
    $N_\mathrm{v}=72$ & 42.88 & 42.56  & \textbf{44.39}  \\
    \end{tabular}
    \caption{PSNR evaluation of \methodname trained with different distillation loss, including $\ell_1$ loss and $\ell_2$ loss commonly used in knowledge distillation, as well as the proposed one with $\mathcal{L}_\mathrm{rdd}$ and $\mathcal{L}_{\mathrm{bcd}}$. All models are trained for 40 epochs considering the computational cost.}
    \label{tab:ablation_disloss}
\end{table}
Table~\ref{tab:ablation_disloss} exhibits the results of \methodname trained with different distillation loss. Our findings suggest that pixel-wise distillation losses, such as $\ell_1$ and $\ell_2$ loss, are not as effective as the proposed one. This is attributed to the fact that conventional distillation tasks involve both the student and teacher networks sharing the same input and ground truth. Consequently, the domain gap does not affect them. However, for sparse-view CT reconstruction, it is arduous for the student to recover the missing information entirely. This renders pixel-wise losses too abrupt for distillation purposes.

\subsection{Ablation on Band-pass-specific Contrastive Distillation}
\begin{figure}[htb] 
    \centering 
    \includegraphics[width=0.45\textwidth]{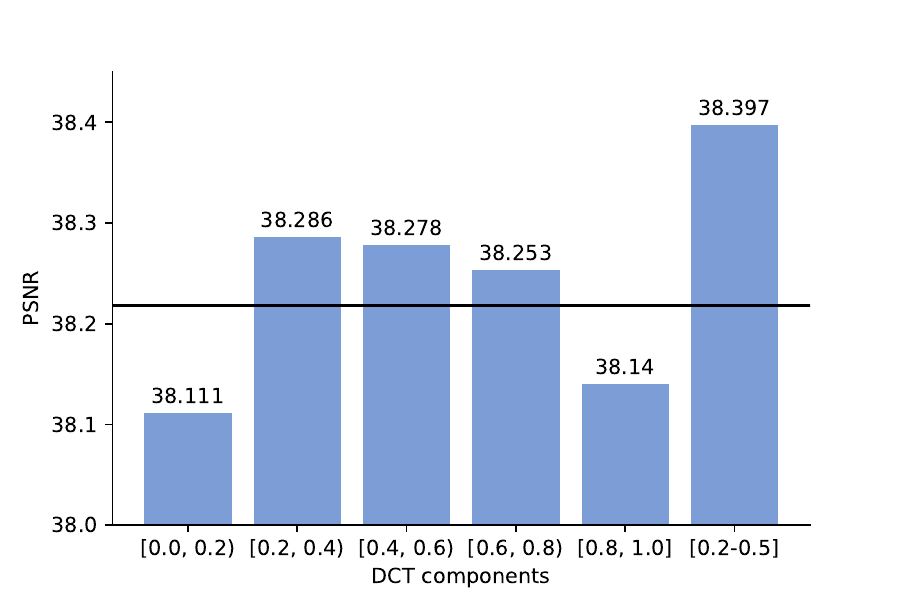}
    \caption{The effect of different frequency components. Note that the black horizontal line represents the contrastive distillation without projecting the representation to the DCT domain. The models are trained with $\mathcal{L}_{\mathrm{bcd}}$ only for 60 epochs.} 
    \label{fig:ablationbpscl} 
\end{figure}

We have demonstrated the effective components by training \methodname with $\mathcal{L}_{\mathrm{bcd}}$ on specific frequency components. However, there are various methods to split the frequency components~\cite{fsdr,fcanet,fda,dctdynamicsr}. Note that in 2D discrete cosine transform, low-frequency components are placed on the upper left. We define the mask $ \mat{M} \in \{0,1\}^{N_\mathrm{w} \times N_\mathrm{h}}$ as follows to select the target components:
\begin{align}
    \mat{M}_{i,j}  =
    \begin{cases}
        1, \text{if}\ i\!\in\![b_\mathrm{low}N_\mathrm{w}, b_\mathrm{up}N_\mathrm{w}] \ \text{and} \ j\!\in\![b_\mathrm{low}N_\mathrm{h},  b_\mathrm{up}N_\mathrm{h}] \\
        0, \text{otherwise,}
        \end{cases}
\end{align}
where $\mat{M}_{i,j}$ is the element in $\mat{M}$ at position $(i,j)$;  $b_\mathrm{low}$ and $b_\mathrm{up}$ denote the hand-craft ratios defining the lower and upper bounds, respectively, which range from $[0, 1]$.
We then split the DCT spectrum into five groups, demarcated by the intervals $[b_\mathrm{low}, b_\mathrm{up}]$, as illustrated in Fig.~\ref{fig:ablationbpscl}. Notably, the model distilled via the vanilla supervised contrastive loss served as the baseline for our comparative analysis and was denoted by the black horizontal line in Fig.~\ref{fig:ablationbpscl}.
Obviously, models trained with frequency components, except for the lowest and highest, perform better than vanilla ones, demonstrating that selecting band-pass components is effective. In addition, middle groups perform relatively better among different groups, demonstrating the effectiveness of the selected band-pass-specific components. Therefore, we select $[b_\mathrm{low}, b_\mathrm{up}] = [0.2, 0.5]$ to train our final models to balance the performance and memory usage.

\section{Efficiency}

\begin{table}[htb]
    \centering
    \tiny
    \begin{tabular}{c|cc|ccc|c}
    Methods &  
    DDNet&
    FBPConvNet&
    DuDoNet&
    DDPTrans& 
    DuDoTrans& 
    \methodname 
    \\
    \shline
    mem. (MB) & 86.4  & 274.9 & 2150.1 & 7220.3 & 3108.5 & 798.8  \\
    infer. (ms) & 14.7  & 11.7 & 49.6 & 71.3 & 78.4 & 33.1  \\
    \end{tabular}
   \caption{Peak memory usage and mean inference time on a single RTX 3090 GPU using 1000 images, with a batch size of 1, at a resolution of $256 \times 256$.} 
   \label{tab:response1_efficiency}
\end{table}

Table~\ref{tab:response1_efficiency} presents the peak memory usage (mem.) and mean inference time (infer.) assessed on a single RTX 3090 GPU with a batch size of 1, averaging over 1000 images at a resolution of $256 \times 256$. Overall, dual-domain methods exhibit lower efficiency compared to image post-processing techniques. Transformer-based methods are suboptimal in both memory usage and inference time to those built with CNN. In contrast, \methodname demonstrates comparable performance to other post-processing methods while achieving higher efficiency than dual-domain approaches by eliminating the need for the teacher network during inference.

%% file: section/more_details.tex
\section{More Visualization Results}

\begin{figure*}[h]
    \centering
    \includegraphics[width=1\linewidth]{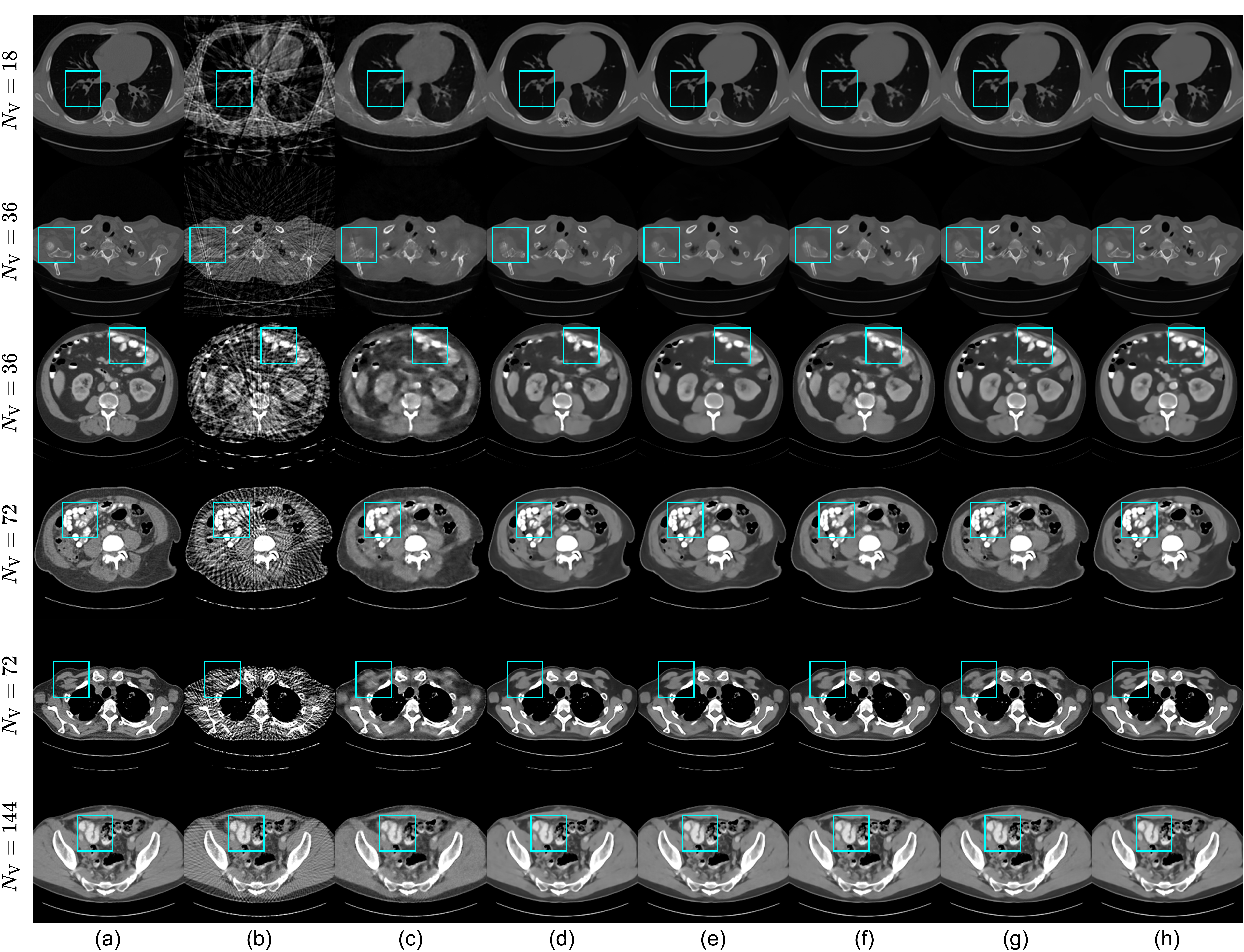}
    \caption{Visual comparison of state-of-the-art methods on DeepLesion dataset: (a) Ground Truth, (b) FBP, (c) DDNet, (d) FBPConvNet, (e) DuDoNet, (f) DDPTrans, (g) DuDoTrans, and (h) \methodname. From top to bottom:  the results under $N_\mathrm{v}=18, 36, 36, 72, 72, 144$;  display window is set to [-1000, 2000] HU for the first and the second rows, and [-200, 300] HU for the rests.}
    \label{fig:deep} 
\end{figure*}

Fig.~\ref{fig:deep} presents the visualization results of six groups of sparse-view images. Among all the methods, \methodname better recovers the clinical details such as the lung trachea in the first row, the round soft tissue in the second row, and the clear boundary highlighted in the fifth row.

\begin{figure*}[h]
    \centering
    \includegraphics[width=1\linewidth]{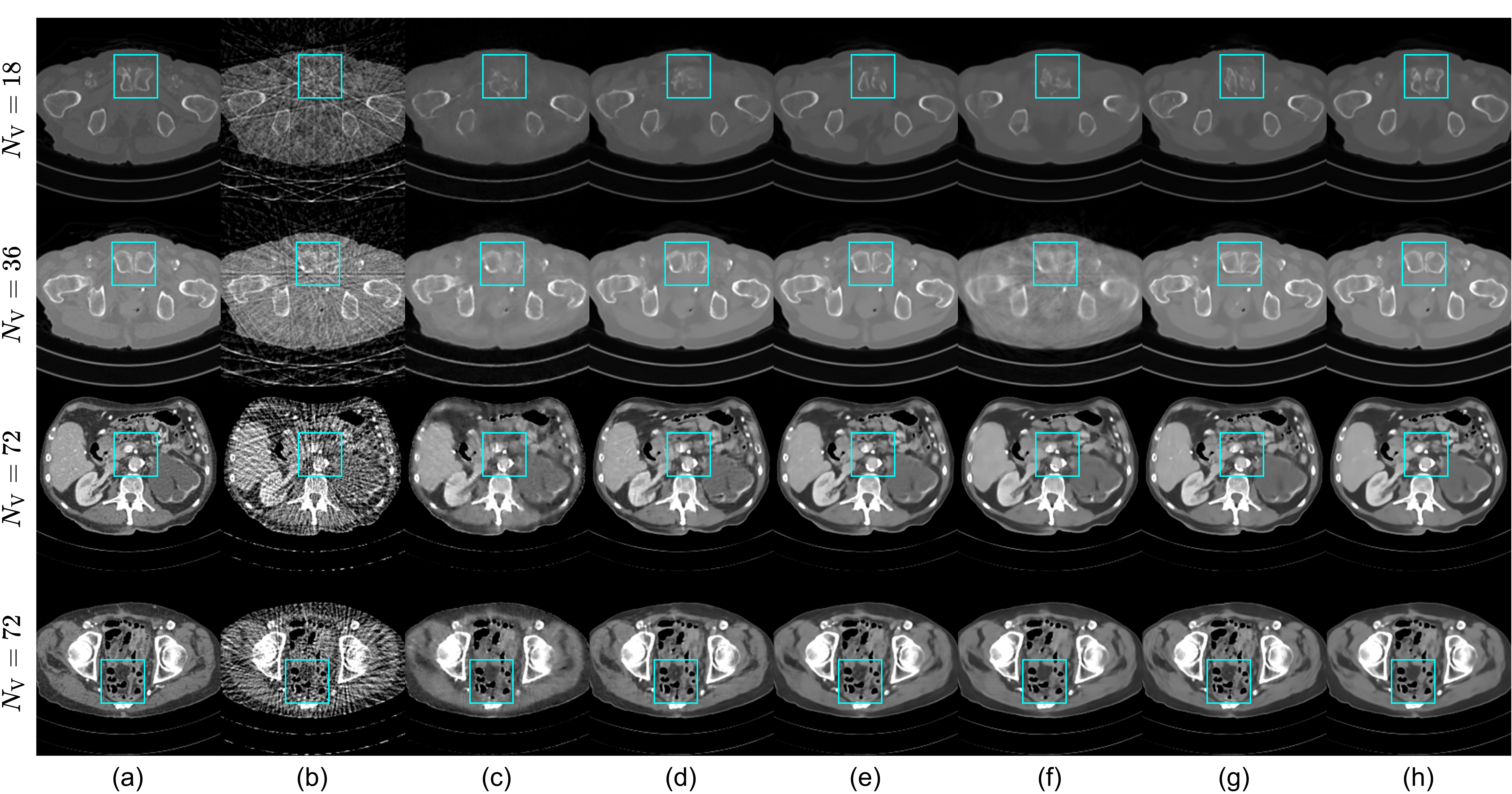}
    \caption{Visual comparison of state-of-the-art methods on AAPM dataset: (a) Ground Truth, (b) FBP, (c) DDNet, (d) FBPConvNet, (e) DuDoNet, (f) DDPTrans, (g) DuDoTrans, and (h) \methodname. From top to bottom:  the results under $N_\mathrm{v}=18, 36, 72, 72$;  display window is set to [-1000, 2000] HU for the first row, [-1000, 1000]HU for the second row and [-200, 300] HU for the third and fourth row.}
    \label{fig:aapm} 
\end{figure*}

Fig.~\ref{fig:aapm} shows another four images in the AAPM dataset. We note that in ultra-sparse scenarios when $N_\mathrm{V} =18, 36$ as shown in the first and the second rows, only \methodname precisely reconstructs the structure highlighted by the blue box. When $N_\mathrm{V}=72$, \methodname achieves competitive performance compared with DuDoTrans but without using the sinogram data.

% \newpage

\section{Detailed Network Architectures}

Tables~\ref{tab:encoder} and~\ref{tab:decoder} show the detailed network architecture of the encoder and decoder, respectively. 

\begin{table*}[!htb]
    \centering
    \renewcommand{\arraystretch}{1.3}{
        \begin{tabular}{lcl} 
            \shline
            \textbf{Name}                         & \textbf{Channels} & \textbf{Description}                      \\ 
            \hline
            Input                        & 2        & sparse-view images $\mat{I}_\mathrm{S}$ or $\mat{I}_\mathrm{T}$ \\                                                                                                                                
            \hline
            Rpad0                        & 2        & reflectionpad2d((3,3,3,3))                                                                                                                                                                                   \\ 
            \hline
            Down1                        & 64       & K7C64S1P1-BN-ReLU                                                                                                                                                                                            \\ 
            \hline
            Down2                        & 128      & K3C128S2P1-BN-ReLU                                                                                                                                                                                           \\ 
            \hline
            \textbf{FFC-Split} $\times m$        & 256      & \begin{tabular}[c]{@{}l@{}}local branch: K3C64S2P1-BN-ReLU \\ global branch: K3C192S2P1-BN-ReLU\end{tabular}                                                                                                 \\ 
            \hline
            \textbf{FFC-1} $\times m$   & 256      & \begin{tabular}[c]{@{}l@{}}convl2l: K3C64S1P1\\ convl2g: K3C192S1P1\\ convg2l: K3C64S1P1\\ convg2g: K1C96S1-bn-relu-FFT-K1C192S1-iFFT-K1C192S1\\ local branch: BN-ReLU\\ global branch: BN-ReLU\end{tabular}  \\ 
            \hline
            \textbf{FFC-2} $\times m$   & 256      & \begin{tabular}[c]{@{}l@{}}convl2l: K3C64S1P1\\ convl2g: K3C192S1P1\\ convg2l: K3C64S1P1\\ convg2g: K1C96S1-bn-relu-FFT-K1C192S1-iFFT-K1C192S1\\local branch: BN-ReLU\\global branch: BN-ReLU\end{tabular}    \\ 
            \hline
            \textbf{FFC-Cat} $\times m$ & 256      & concat(local branch, global branch) w/ residual learning                  \\
            \shline

    \end{tabular}}
    \vspace{5pt}
    \caption{Network architecture of student and teacher encoder.  We use `\texttt{K-C-S-P}' to denote the kernel, channel, stride, and padding configuration of convolution layers.} 
    \label{tab:encoder}
\end{table*}

\begin{table*}[!htb]
    \centering
    \renewcommand{\arraystretch}{1.3}{
        \begin{tabular}{lcl} 
            \shline
            \textbf{Name}                         & \textbf{Channels} & \textbf{Description}                      \\ 
            \hline

            \textbf{FFC-Split} $\times n$       & 256      & \begin{tabular}[c]{@{}l@{}}local branch: K3C64S2P1-BN-ReLU \\ global branch: K3C192S2P1-BN-ReLU\end{tabular}             \\ 
            \hline
            \textbf{FFC-1} $\times n$   & 256      & \begin{tabular}[c]{@{}l@{}}convl2l: K3C64S1P1\\ convl2g: K3C192S1P1\\ convg2l: K3C64S1P1\\ convg2g: K1C96S1-bn-relu-FFT-K1C192S1-iFFT-K1C192S1\\ local branch: BN-ReLU\\ global branch: BN-ReLU\end{tabular}  \\ 
            \hline
            \textbf{FFC-2} $\times n$   & 256      & \begin{tabular}[c]{@{}l@{}}convl2l: K3C64S1P1\\ convl2g: K3C192S1P1\\ convg2l: K3C64S1P1\\ convg2g: K1C96S1-bn-relu-FFT-K1C192S1-iFFT-K1C192S1\\local branch: BN-ReLU\\global branch: BN-ReLU\end{tabular}    \\ 
            \hline
            \textbf{FFC-Cat} $\times n$ & 256      & concat(local branch, global branch) w/ residual learning                  \\
            \hline
            Up1                          & 128      & ConvTranspose2d: K3C128S2P1-BN-ReLU                                                                                                                                                                          \\ 
            \hline
            Up2                          & 64       & ConvTranspose2d: K3C64S2P1-BN-ReLU                                                                                                                                                                           \\ 
            \hline
            Rpad1                        & 64       & reflectionpad2d((3,3,3,3))                                                                                                                                                                                   \\ 
            \hline
            Out                          & 1        & K7C1S1     \\
            \shline

    \end{tabular}}
    \vspace{5pt}
    \caption{Network architecture of the shared decoder. We use `\texttt{K-C-S-P}' to denote the kernel, channel, stride, and padding configuration of convolution layers.} 
    \label{tab:decoder}
\end{table*}
\clearpage
\clearpage